\documentclass[aps,prl,superscriptaddress,floatfix,a4,showpacs,twocolumn,nofootinbib]{revtex4}

\newcommand{\be}{\begin{eqnarray}}
\newcommand{\ee}{\end{eqnarray}}

\def\hbar#1{\slash\hspace{-2.5mm}#1}

\usepackage{graphicx}
\usepackage{dcolumn}
\usepackage{bm}

\def\be{\begin{equation}}
\def\ee{\end{equation}}
\def\bea{\begin{eqnarray}}
\def\eea{\end{eqnarray}}

\def\lsim{\raise0.3ex\hbox{$\;<$\kern-0.75em\raise-1.1ex\hbox{$\sim\;$}}}
\def\gsim{\raise0.3ex\hbox{$\;>$\kern-0.75em\raise-1.1ex\hbox{$\sim\;$}}}

\begin{document}
\title{Gauged $B-L$ Leptogenesis}

\author{Y. Kajiyama}
\affiliation{National Institute of Chemical Physics and
Biophysics, Ravala 10, Tallinn 10143, Estonia}
\author{S. Khalil}
\affiliation{Centre for Theoretical Physics, The British
University in Egypt, El Sherouk City, Postal
No. 11837, P.O. Box 43, Egypt.} %
\affiliation{Department of Mathematics, Ain Shams University,
Faculty of Science, Cairo, 11566, Egypt.}
\author{E. Ma}
\affiliation{Physics and Astronomy Department, University of
California, Riverside, CA 92521, USA} %
\author{H. Okada}
\affiliation{Centre for Theoretical Physics, The British
University in Egypt, El Sherouk City, Postal No. 11837, P.O. Box
43, Egypt.}
\date{\today }

\begin{abstract}
We propose a new leptogenesis scenario in a gauged $B-L$ model
with supersymmetry at the TeV energy scale.  Instead of relying on
the very small Yukawa couplings of the singlet neutrinos $N^c$ to
generate the observed baryon asymmetry of the Universe, which
requires a very large resonance enhancement, their $B-L$ gauge
interactions are invoked.  Successful leptogenesis is then
possible if a particular scalar bilinear $\tilde{N}^c \tilde{N}^c$
term is disallowed.
\end{abstract}

\pacs{12.60.Cn; 11.30.Fs. }


\maketitle
The current measurement of the baryon-to-entropy ratio of the Universe is given by \cite{Jungman:1995bz} %
\be%
Y_B\equiv \frac{n_B}{s} = (0.87 \pm 0.02) \times 10^{-10}, %
\label{ybexp}
\ee %
where $s = 2\pi^2 g_{\star} T^3/45$ is the entropy density and
$g_\star$ is the effective number of relativistic degrees of
freedom. CP violation is an essential requirement in order to
obtain this asymmetry. Leptogenesis \cite{yanagida} is the most promising mechanism to explain it.
It is known that there are several scenarios of the leptogenesis \cite{resonance,hierarchy, Abbas:2007ag,babu}.

Leptogenesis through the decay of a heavy singlet neutrino is
considered as the best scenario for understanding the observed
baryon asymmetry of the Universe.  However, the energy scale
involved in a successful application is usually in the range
$10^9$ to $10^{13}$ GeV, which renders the idea impossible to
verify experimentally.  It has also been suggested that this
mechanism works just as well at the more easily accessible TeV
energy scale, but then the very small Yukawa couplings required by
neutrino masses implies that this effect is much too small to be
viable, unless it is compensated by a very large resonance
enhancement \cite{resonance}, i.e. the near mass degeneracy of two singlet
neutrinos.  As an alternative solution, instead of the
resonance-enhancement hypothesis, we suggest that the source of
this matter-antimatter asymmetry is actually a gaugino interaction of 
gauged $U(1)_{B-L}$
symmetry in a supersymmetric extension of the Standard Model (SM)
at the TeV scale.

In supersymmetry, the addition of the singlet superfield
$\hat{N}^c$ with $B-L=1$ implies a fermion $N^c$ and a scalar
$\tilde{N}^c$.  As $B-L$ is spontaneously broken by singlet
superfields $\hat{\chi}_{1,2}$ with $B-L=\mp 2$, an exact $Z_2$
residual symmetry remains, i.e. $R$ parity, with $R =
(-)^{3(B-L)+2j}$.  As a result, $N^c$ acquires a large Majorana mass
through $\langle \chi_1 \rangle N^c N^c$, so that it may decay into
both leptons and antileptons, thereby initiating leptogenesis.  As
for $\tilde{N}^c$, there are in general two kinds of mass terms:
$(\tilde{N}^c)^* \tilde{N}^c$ and $\tilde{N}^c \tilde{N}^c$.  If
the latter is absent, then $\tilde{N}^c$ may be assigned $L=-1$ in
a subset of its interactions.  This will be the key to having a
successful leptogenesis scenario, using $B-L$ gauge interactions. 
Some of us have studied resonant scenarios in {\rm TeV} scale $B-L$ model in Ref.\cite{Abbas:2007ag}.

Consider the two families $N^c_{1,3}$ and $\tilde{N}^c_{1,3}$ with masses
arranged in the order
\begin{equation}
M_{\tilde{N}^c_3} < M_{{N}^c_1} < M_{{N}^c_3} < M_{\tilde{N}^c_1},
\end{equation}
and with $N^c_1$ coupling to $g_{B-L} \tilde{Z}_{B-L} (\tilde{N}^c_1 \cos \tilde{\theta}
+ \tilde{N}^c_3 \sin \tilde{\theta})$ and $N^c_3$ coupling to the orthogonal
combination, where $g_{B-L}$ is the $B-L$ gauge coupling, and $\tilde{Z}_{B-L}$ is
the $B-L$ gaugino which is also assumed to be lighter than the mass difference
between $N^c_1$ and $\tilde{N}^c_3$.  The decay of $N^c_1$ is then only
into $\tilde{N}^c_3 + \tilde{Z}_{B-L}$, with coupling $g_{B-L} \sin \tilde{\theta}$.
Since $\tilde{\theta}$ represents the misalignment of the two families after
supersymmetry breaking, it may be assumed to be very small, i.e. of order
$10^{-6}$, to satisfy the out-of-equilibrium condition for $M_{N^c_1}$ at the
TeV scale.  A large lepton asymmetry proportional to
$(g_{B-L} \cos \tilde{\theta})^2$
may then be generated through the one-loop exchange of $N^c_3$, provided
that below $M_{N^c_1}$, additive lepton number is conserved, i.e.
$\tilde{N}^c_3$ having $L=-1$ in all its subsequent interactions.
In the following we will show in detail how this all works.

As shown in Ref.\cite{Khalil:2007dr}, after the $B-L$ symmetry breaking by the
VEVs $\langle \chi_{1,2} \rangle=v'_{1,2}$ (we define $v'_1=v'\sin
\theta$ and $v'_2=v' \cos \theta$), a bilinear coupling $B_{Nij}^2
\tilde N^c_i \tilde N^c_j$ is generally obtained and it is given
by $B_N^2=-v'_1Y_N^A+Y_Nv'_2 \mu'^*$. 
Here, we assume that
$B_N^2=0$ so that the off-diagonal elements of sneutrino
$\tilde{N}^c$ mass matrix, in the $(\tilde N^c_i, \tilde
N^{c*}_i)$ basis, vanish. Therefore, $\tilde N^c$ and $\tilde
N^{c*}$ are mass eigenstates with mass squared $M_N^* M_N^T+\tilde
m_{\tilde N^c}^2 +\frac{1}{4}M_{Z_{B-L}}^2\cos 2\theta$, and they
have lepton numbers $L = \mp 1$ respectively. Moreover, if $\cos 2
\theta$ is negative, $\tilde N^c$ can be lighter than $N^c$.
Actually, only one $\tilde{N}^c$ mass eigenstate needs to have
lepton number and be lighter than the lightest $N^c$.  This is the
crucial assumption of our proposal. For $B-L$ neutralinos $\tilde
\chi_a=(\tilde \chi_1, \tilde \chi_2,-i \tilde Z_{B-L})$
\cite{Khalil:2008ps}, the mass eigenstates ${{\tilde\chi_{phy.a}}}
(a=1,2,3)$ are given by the unitary diagonalization matrix $R$ as
$\tilde \chi_a=\sum_bR_{ab}\tilde \chi_{phy.b},~~R^{\dag}R=1$. In
our numerical calculation, we derive mass eigenvalues and mixing
matrix $R$ in the following two limiting cases: Case A)~ $\mu' , ~M_{Z_{B-L}}\gg M_{B-L}$,
Case B)~$ M_{B-L},~M_{Z_{B-L}}\gg \mu'$,
where $\mu'$, $M_{B-L}$, and $M_{Z_{B-L}}$ are defined as the mass parameter of $\hat\chi_{1,2}$, 
${\widetilde Z_{B-L}}$, and $Z_{B-L}$, respectively. 

The Lagrangian, in flavor eigenstates, relevant
for our analysis is given by %
\bea %
{\cal L}&=& -\sqrt{2}g_{B-L} (-i \tilde Z_{B-L})\tilde
N_i^{c*} (N_i^c)-Y_{Nij}\tilde \chi_{1} (N_i^c)\tilde N_j^{c}
\nonumber \\
&-&M^2_{\tilde N^c ij}\tilde N_i^{c*}\tilde N_j^{c}-\frac{1}{2}
M_{Nij}(N_i^c) (N_j^c)+h.c., \label{lag} %
\eea %
where %
\bea%
M_{Nij}&=&Y_{Nij}v' \sin \theta,
\label{Nmass}\\
M^2_{\tilde N^c ij}\!&\!=\!&\!(M_N^* M_N^T)_{ij}\!+\!m_{\tilde N^c
ij}^2\!+\!\frac{1}{4}M_{Z_{B-L}}^2 \cos 2 \theta\delta_{ij}.
\label{Ntildemass} %
\eea%
These mass matrices are diagonalized by unitary matrices $U$ and
$\Gamma$: $ U^T M_N U={\rm diag},~\Gamma^{\dag} M_{\tilde N^c}^2
\Gamma={\rm diag}$, and mass eigenstates $(N^c)_L^m$ and $(\tilde
N^c)^m$ are defined as $(N_i^c)=U_{ij} (N^c_j)^m,~~ (\tilde
N^c_i)=\Gamma_{ij}(\tilde N^c_j)^m$.  Notice that the mixing
matrix $U$ and $\Gamma $ are in general different from each other.
Therefore, the combination $U^{\dag}\Gamma$ is not unit matrix,
and complex. 
This is the origin of CP violation.

The Lagrangian in mass eigenstate (hereafter we remove  the
index ``$m$") is given by %
\bea %
{\cal L}&=&-A_{aij}\bar \Psi_a P_L N_i \tilde N^{c*}_j-B_{aij}\bar \Psi_a P_L N_i \tilde N^c_j+h.c.\nonumber \\
&-&M_{\tilde N^c i}^2\tilde N^{c*}_i\tilde N^{c}_i
-\frac{1}{2}M_{Ni}\bar N_i N_i, \label{lag2} %
\eea %
where %
\be %
A_{aij}=\sqrt{2}g_{B-L}R_{3a}(\Gamma^{\dag}U)_{ji},~
B_{aij}=Y_{Ni}R_{1a}(U^{\dag}\Gamma)_{ij}. %
\ee%
The four-component Majorana spinors are defined as
$\Psi_a=(\tilde \chi_{phy.a},\bar{\tilde{\chi}}_{phy.a})^T$ and 
$N_i=(N^c_i,\bar{N}^c_i)^T$. 
Notice that $A,B ={\cal O}(1)$ naturally for $i=j$ and small for $i
\neq j$ as shown below Eq. (\ref{epapprox}).

Now, we consider leptogenesis by $N_1 \to \Psi_a \tilde
N^c_j$ induced by Eq. (\ref{lag2}), assuming the mass hierarchy
$M_{N1}\ll M_{N2,3}$. We assume that only the lightest $B-L$
neutralino $\Psi \equiv \Psi_1$ and sneutrino of the third
generation $\tilde N^c_3 $ are lighter than $N_1$, and satisfy the
relation $m_{\tilde \chi}+M_{\tilde N^c_3}<M_{N1}$. Moreover, we
restrict ourselves to two limiting cases. 
Since $R_{11} (R_{13})\ll 1$ in the case A (B),
$A_{aij}(B_{aij}),~(a=1,i=1,j=3)$ gives dominant contributions, and 
$\Psi$ is nearly $\tilde Z_{B-L}(\tilde \chi_1)$. 
As emphasized, due to the fact that $B_N=0$, $\tilde
N^c_3$ carries lepton number, hence the decay $N_1 \to
\Psi \tilde N^c_3$ violates lepton number.

\begin{figure}[t]
\unitlength=1mm
\begin{picture}(170,18)
\includegraphics[width=8cm]{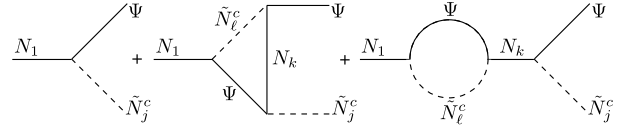}
\end{picture}
\vspace{-0.5cm} \caption{Tree and one-loop diagram of $N_1 \to
\Psi \tilde N_j^c$ decay.} \label{diag}
\end{figure}

CP asymmetry of $N_1 \to \Psi \tilde N^c_3$ decay processes is
generated by the interference between tree and one-loop level
diagrams of vertex and self-energy correction shown in FIG.
\ref{diag}. It is defined as%
\be %
\epsilon_1=\frac{ \Gamma(N_1 \to
\Psi  \tilde N_3^c)- \Gamma(N_1 \to \Psi  \bar{\tilde{N}}^c_3)}{
\Gamma(N_1 \to \Psi \tilde N_3^c)+ \Gamma(N_1 \to \Psi \bar{
\tilde N}_3^c)}. \label{ep} %
\ee %
The decay rate $\Gamma$ at
one-loop level is given by %
\bea %
\!\!\!\!\Gamma(N_1 \!\to\! \Psi \tilde
N_3^c)\!\!&\!\!=\!\!&\!\!\frac{1}{2M_{N1}}\!\left|\! {\cal
A}_{tree}\!\!+\!\!{\cal A}_{loop}F\!\right|^2 \!\!I_2(\!N_1 \!\to\! \Psi
\tilde N_3^c\!),~
\label{d1}\\
\!\!\!\!\Gamma(N_1 \!\to\! \Psi \bar {\tilde
N}_3^c)\!\!&\!\!=\!\!&\!\!\frac{1}{2M_{N1}}\!\left|\! {\cal
A}^*_{tree}\!\!+\!\!{\cal  A}^*_{loop}F\!\right|^2 \!\!I_2 (\!N_1
\!\to \!\Psi
\!\tilde N_3^c\!),~ \label{d2} %
\eea%
where the phase space integral of two-body decay $I_2$ is given by
$$
\!\!\!I_2(\!X\!\!\to\!\! YZ)\!=\!\frac{1}{\!8\pi\!
M_X^2\!}\sqrt{\![\!M_X^2\!-\!(\!M_Y\!+\!M_Z\!)^2\!][\!M_X^2\!-\!(M_Y\!-\!M_Z\!)^2\!]\!}~.
$$
${\cal A}_{tree,loop}$ are tree and loop level amplitudes, and $F$ is
kinematical factor. As the loop-level diagrams have vertex and
self-energy corrections, we write ${\cal A}_{loop}F$ as:
${\cal A}_{loop}F={\cal A}_{v}F_v+{\cal A}_s F_s$. From Eqs.
(\ref{d1}) and (\ref{d2}), the total CP asymmetry
$\epsilon_1=\epsilon_1^v+\epsilon_1^s$ is given by %
\be
\epsilon_1^{v(s)}=-\frac{2 {\rm Im}\left[ {\cal A}_{tree}^* {\cal A}_{v(s)}\right]{\rm Im}\left[F_{v(s)}\right]}
{\left|{\cal A}_{tree}\right|^2}, 
\ee
with
\begin{eqnarray*}
\left|{\cal A}_{tree}\right|^2&=&M_{N1}^2 (1+r_{\tilde \chi}-r_{\tilde N_3})\left| A_{113}\right|^2,\\
{\cal A}_{tree}^* {\cal A}_v&=&\sum_k M_{N1}M_{Nk} \left(A_{113}^*A_{1k3}\right)^2,\\
{\cal A}_{tree}^* {\cal A}_s&=&\sum_k M_{N1}M_{Nk}
\left(A_{113}^*A_{1k3}\right)^2
\frac{1+r_{\tilde \chi}-r_{\tilde N3}}{1-r_{Nk}},\nonumber\\
F_v=\frac{1}{(4 \pi)^2}&&\!\!\!\!\!\!\!\!\!\!\!\!\int dx dy
dz\delta(x+y+z-1)\nonumber\\
\times&&\!\!\!\!\!\!\!\!\!\!\!\!\frac{(y-1)(1+r_{\tilde \chi}-r_{\tilde N3})
+2 z r_{\tilde \chi}}{-xy-yzr_{\tilde N3}-zxr_{\tilde \chi}+x r_{\tilde N3}+yr_{\tilde \chi}+z r_{Nk}},\\
F_s=\frac{1}{(4 \pi)^2}&&\!\!\!\!\!\!\!\!\!\!\!\!\int dx dy\delta(x+y-1)\ln (-xy+x
r_{\tilde N3}+yr_{\tilde \chi}), \nonumber%
\end{eqnarray*}
for the case A, and $A\to B^*$ for the case B, 
where $r_X=m_X^2/M_{N1}^2$. 
In both cases, CP asymmetry $\epsilon_1$ of $N_1 \to \Psi \tilde
N^c_3$ decay has the structure %
\be%
\epsilon_1 \sim \sum_k
\frac{{\rm Im}\left[\Gamma_{13} \Gamma_{13}\Gamma_{k3}^*
\Gamma_{k3}^*\right]}{|\Gamma_{13}|^2},%
\label{epapprox} %
\ee%
where $U=1$ is assumed. From Eq. (\ref{epapprox}), one finds
$\Gamma_{13}\sim\sin\tilde{\theta}\ll1$, as required by out-of equilibrium condition:
$\Gamma (N_1 \to \Psi \tilde N_3^c)< H(z=1)$, for $z=M_{N1}/T$. Also,
$\Gamma_{33}\sim\cos\tilde{\theta}\sim 1$, which leads to a large CP
asymmetry. This situation is realized if the mixing matrix
$\Gamma$ is almost diagonal.

In our model, baryon asymmetry is obtained through the following procedure:
\begin{enumerate}
\item $N_1^c\to \Psi \tilde N_3^c$ decay generates $\tilde N_3^c$ asymmetry $Y_{\Delta \tilde N_c}$. 
\item $\tilde N_3^c$ decays into (s)lepton by Dirac Yukawa couplings, soft SUSY 
breaking A-term and $\mu$-term, and resulting (s)lepton asymmetry $Y_{\Delta L(\Delta \tilde L)}$ 
is obtained by solving the Boltzmann equations. 
\item Sphaleron
converts total lepton asymmetry $Y_L=Y_{\Delta L}+Y_{\Delta \tilde L}$ 
to baryon asymmetry $Y_B$.
\end{enumerate}
Moreover, we take into account
scattering processes mediated by $B-L$ gauge boson : 
$N_1N_1\to Z_{B-L}\to f\bar f$. For the elastic scattering ($f=N_1$), 
the scattering rate is large for high temperature $z\ll1$, which realizes kinetic equilibrium. 
For very large $M_{Z_{B-L}}$, the scattering is
Boltzmann suppressed near $z=1$. So decay dominates at this
temperature, and leptogenesis occurs at $z=1$. 
This condition may give lower bound as $M_{Z_{B-L}}\gsim 10^3 M_{N1}$ 
\cite{lutyb-l,Buchmuller:1997nz,masu2r}.
On the other hand
for small $M_{Z_{B-L}}$ case, scattering contributions survive
until $z\sim10$. 
As a result, since asymmetry due to the decay starts to be produced by small $N_1$ abundance 
at large $z$, 
only small lepton asymmetry is created unless CP asymmetry is large. 
On the other hand, since scattering processes by $B-L$ gaugino; 
$N_1\tilde N_3^c \to \tilde Z_{B-L} \to f \bar {\tilde f}$, are well
suppressed by small mixing matrix $\Gamma_{13}\ll 1$, we neglect them.

The thermal average decay and scattering rates that contribute to Boltzmann equations, which we
solved numerically to get the total lepton asymmetry, are give by:
\bea%
\gamma_D&=&n_{N1}^{eq}\frac{K_1(z)}{K_2(z)}
\left[ \Gamma(N_1\to \tilde N_3^c \Psi)+\Gamma(N_1\to \bar {\tilde N}_3^c \Psi)\right],\nonumber\\
\gamma_{\tilde N^c \bar L}&=&n_{\tilde N_3^c}^{eq}\frac{K_1(\sqrt{r_{\tilde N_3}}z)}
{K_2(\sqrt{r_{\tilde N_3}}z)}\Gamma(\tilde N_3^c \to \bar L\bar{\tilde H}_2),\nonumber\\
\gamma_{\tilde N^c \bar{\tilde L}}&=&n_{\tilde
N_3^c}^{eq}\frac{K_1(\sqrt{r_{\tilde N_3}}z)}{K_2(\sqrt{r_{\tilde
N_3}}z)}
\left[\Gamma(\tilde N_3^c \to \bar{\tilde L}\bar{H}_2)+\Gamma(\tilde N_3^c \to \bar{\tilde L}{H}_1)\right],
\nonumber\\
\gamma_{\tilde N^c {\tilde L}}\!\!&\!=\!&\!\!n_{\!\tilde N_3^c}^{eq}\!\frac{K_1(\sqrt{r_{\tilde N_3}}z\!)}{\!K_2(\sqrt{r_{\tilde
N_3}}z\!)}
\!\Gamma(\!\tilde N_3^c \!\to\! {\tilde L}{H}_2\!),\nonumber\\
\gamma_S&=&\langle \sigma \rangle=\frac{T}{64 \pi^4} 
\int_{s_{max}}^{\infty}ds \sqrt{s}~\hat \sigma(s)K_1\left(\frac{\sqrt{s}}{T}\right),
\eea%
where $K_1(z)$ and $K_2(z)$ are modified Bessel functions, and 
$s_{max}={\rm max}[4M_{N1}^2,4m_{f}^2]$.
The decay rate of $\tilde{N}^c_3$ at $T=0$ into $\bar{L} \bar{\tilde H}_2$,
$\bar{\tilde L}\bar{H}_2$,$\bar{\tilde L}{H}_1$, and ${\tilde L}{H}_2$ is written as %
\be %
\!\!\!\!\Gamma(\tilde N_3^c \to AB)
=\sum_i\frac{1}{M_{\tilde N^c_3}}\left| \left( Y_{AB}\right)_{3i}\right|^2 I_2(\tilde N_3^c \to A B),
\ee%
where the associated couplings $Y_{AB}$ are given by $ \left(\Gamma^T Y_{\nu}\right)_{3i} (M_{\tilde N^c_3}^2-m_{\tilde H_2}^2)^{1/2}$, $\left(\Gamma^TA_{\nu}\right)_{3i}$, $\mu^* \left( \Gamma^T
Y_{\nu}\right)_{3i}$, and $\left(\Gamma^{\dag}M_NY_{\nu}\right)_{3i}$.
The reduced cross section $\hat \sigma(s)$ for fermionic (bosonic) final states $N_1N_1 \to Z_{B-L}\to
\psi\bar \psi (\phi \bar \phi)$ 
is given by\footnote{In Ref. \cite{Buchmuller:1997nz},  
$\langle \sigma \rangle$ under approximation of all final states to be massless is given.} %
\bea %
\hat \sigma(s)
&=&
\frac{g_{B-L}^4}{3 \pi}\frac{1}{\left(s-M_{Z_{B-L}}^2\right)^2+
M_{Z_{B-L}}^2 \Gamma_{Z_{B-L}}^2}\sqrt{1-\frac{4M_{N1}^2}{s}}\nonumber\\
\times
&&\!\!\!\!\!\!\!\!\!\!\!\!\left[Q_{\psi}^2\sqrt{1-\frac{4m_{\psi}^2}{s}}
\left[s^2-(4M_{N1}^2+3m_{\psi}^2)s+10M_{N1}^2m_{\psi}^2\right]\right.\nonumber\\
&+&\left.Q_{\phi}^2\sqrt{1-\frac{4m_{\phi}^2}{s}}(s-4M_{N1}^2)(s-4m_{\phi}^2)
\right],
\eea%
where $Q_{\psi,\phi}$ is $U(1)_{B-L}$ charge of the field $\psi$
and $\phi$.


Now we give numerical examples for 
$v'=6~{\rm TeV}$, $ M_{N}= (5,5.5,6)~{\rm TeV}$, 
$M_{\tilde N^c}= (5.7,6.1,0.3)~{\rm TeV}$, $g_{B-L}=1$, $M_{Z_{B-L}}=2\sqrt{2}g_{B-L}v'\simeq 17~{\rm
TeV}$, $g_*=251.25$. 
We focus on the above two cases A and B with the following inputs:
\begin{eqnarray*}
&&({\rm A})~:~
M_{B-L}=300~{\rm GeV},~ \theta=\pi/2,~\mu'=0.9~M_{Z_{B-L}}, \\
&&({\rm B})~:~
M_{B-L}=1.2~M_{Z_{B-L}},~\theta=\pi/4, ~\mu'=300~{\rm GeV}, 
\end{eqnarray*} 
Since $Y_N=M_N/(v' \sin\theta)$ is diagonal, $U=1$. For both cases, 
the scalar mass matrix $m^2_{\tilde N^c}$ of soft SUSY
breaking terms has small deviation from the diagonal form,
which gives small $\Gamma_{13}$. 
In order to obtain light $\tilde N^c_3$, $(m_{\tilde N^c}^2)_{33}$ is tuned
to be $(6.0~{\rm TeV})^2$. 
The corresponding CP asymmetry
$\epsilon_1$ and the out-of equilibrium condition
$\Gamma/H(z=1)$ are given by%
\bea
({\rm A})~:~\epsilon_1&=&-0.10,~\frac{\Gamma}{H(z=1)}=20.0,\\
({\rm B})~:~\epsilon_1&=&-0.080,~\frac{\Gamma}{H(z=1)}=12.4.
\eea%

\begin{figure}[t]
\unitlength=1mm
\begin{picture}(70,40)
\includegraphics[width=6cm]{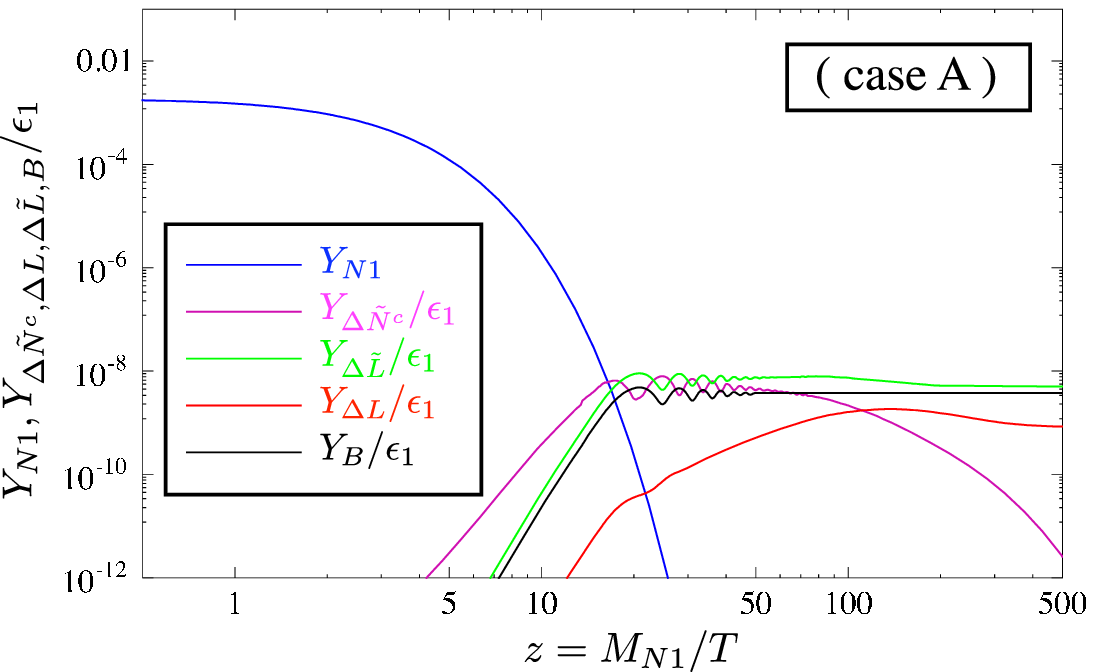}
\end{picture}
\begin{picture}(70,40)
\includegraphics[width=6cm]{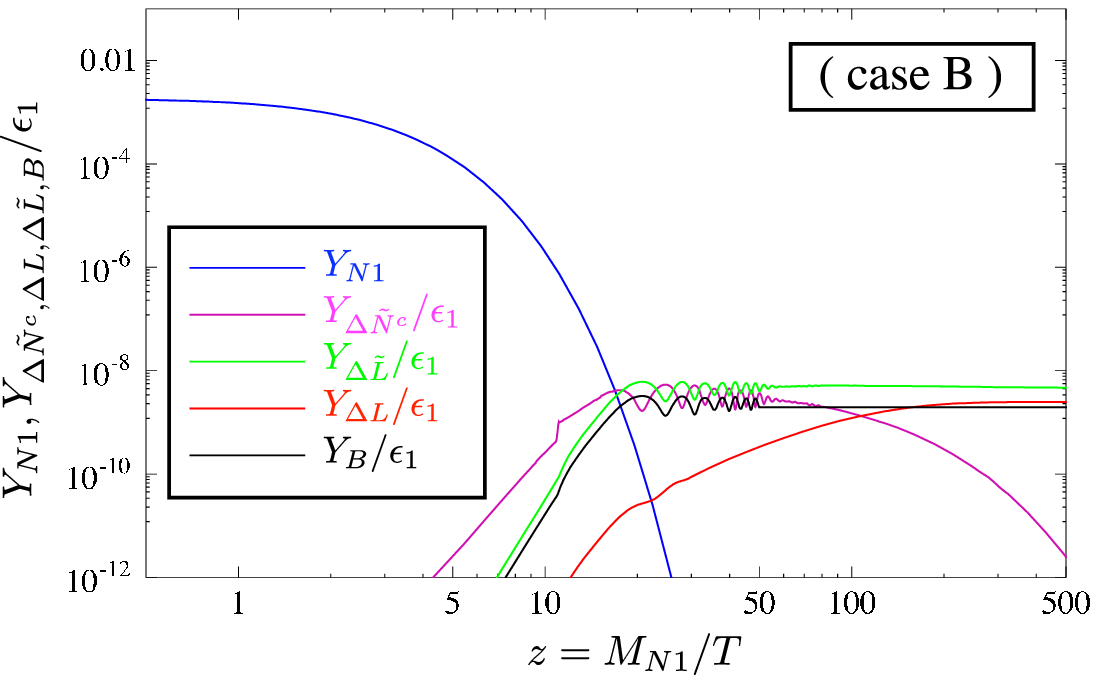}
\end{picture}
\caption{$Y_{N1}$ and $Y_{\Delta \tilde N^c,\Delta L, \Delta \tilde L,B}/\epsilon_1$ for
the case A and B.}
\label{Ys}
\end{figure}

 FIG. \ref{Ys} show the behavior of $Y_{N1}$ and $Y_{\Delta
\tilde N^c,\Delta L,\Delta \tilde L,B}/\epsilon_1$ for the case A
and B. Sphaleron processes \cite{Kuzmin:1985mm} are in equilibrium above
the critical temperature $T_c$. 
In this region, lepton asymmetry is converted into
baryon asymmetry with the rate $Y_B=-8/15 Y_L$. Below $T_c$,
sphaleron processes are still in equilibrium and the conversion
rate from lepton to baryon asymmetry is a function of the temperature-dependent 
VEV $v(T)$ \cite{sphaleron}. At some temperature $T_d<T_c$,
sphaleron processes are switched off due to the Boltzmann factor
and baryon asymmetry never evolves below $T_d$ while lepton
asymmetry still evolves by the Boltzmann equations. However, we
make approximation that sphaleron processes are active for
$T>100~{\rm GeV}$, and switched off for $T<100~{\rm GeV}$.  From
this approximation, we obtain the final results with $Y_{\nu}=3\times 10^{-8}$: %
\bea%
({\rm A})~:~Y_B&=&3.9\times 10^{-10},\\
({\rm B})~:~Y_B&=&1.6\times 10^{-10}.
\eea%
Therefore we can obtain enough baryon
asymmetry.

%
In conclusion, we have shown that a successful TeV scale leptogenesis can take place in gauged $B-L$ supersymmetric model. In this model, if the right-sneutrino bilinear term is absent, then the lightest sneutrino is assigned a lepton number. Therefore if $\tilde N_3^c$ is lighter than $N_1^c$ and 
scalar mass matrix of $\tilde N^c$ is almost diagonal, 
a large lepton asymmetry can be generated 
by $B-L$ neutralino interactions of ${\cal O}(1)$ couplings $g_{B-L}$ and/or $Y_N$ through the one-loop exchange of $N_3^c$ for the decay $N_1^c \to \tilde{N}_3^c \Psi$. This asymmetry of $\tilde{N}_3^c$ is transmitted into asymmetry of lepton and slepton through the Yukawa coupling, trilinear coupling, and $\mu$-term, and sphaleron converts 
lepton asymmetry to baryon asymmetry. Although very heavy $B-L$ gauge boson 
$M_{Z_{B-L}}\gsim 10^3 M_{N1}$ is required for suppress scattering effects in many cases, 
$M_{Z_{B-L}}\sim 3 M_{N1}$ is possible in our model because CP asymmetry is large, $\epsilon_1 \sim 0.1$. 

\vspace{0.2cm}
\begin{center}
{\bf Acknowledgements}
\end{center}
 The work of Y. K. was supported by the ESF grant No. 8090. 
 This work of E. M. was supported in part by the
U.~S.~Department of Energy under Grant No.~DE-FG03-94ER40837.
 S. K. and H. O. acknowledge partial support from the Science
 and Technology Development Fund (STDF) project ID 437 and the ICTP project ID 30.
\bibliographystyle{unsrt}

\end{document}